\documentclass[preprint,journal]{vgtc}       

\ifpdf
  \pdfoutput=1\relax                   
  \usepackage{graphicx}                
  \DeclareGraphicsExtensions{.pdf,.png,.jpg,.jpeg} 
\else
  \usepackage{graphicx}                
  \DeclareGraphicsExtensions{.eps}     
\fi%

\graphicspath{{figures/}{pictures/}{images/}{./}} 

\usepackage{microtype}                 
\PassOptionsToPackage{warn}{textcomp}  
\usepackage{textcomp}                  
\usepackage{mathptmx}                  
\usepackage{times}                     
\usepackage{cite}                      
\usepackage{tabu}                      
\usepackage{booktabs}                  
\usepackage{amsmath}

\onlineid{0}

\vgtccategory{Research}
\vgtcpapertype{application}

\title{Visual Analysis and Detection of Contrails\\ in Aircraft Engine Simulations}

\author{Nafiul Nipu, Carla Floricel, Negar Naghashzadeh, Roberto Paoli, G. Elisabeta Marai %
}
\authorfooter{
Nafiul Nipu, Carla Floricel, Negar Naghashzadeh, Roberto Paoli, and G. Elisabeta Marai are at the University of Illinois at Chicago. E-mail: {mnipu2@uic.edu}.
}

\shortauthortitle{Nipu \MakeLowercase{\textit{et al.}}: Visual Analysis and Detection of Contrails in Aircraft Engine Simulations}

\abstract{Contrails are condensation trails generated from emitted particles by aircraft engines, which perturb Earth's radiation budget. Simulation modeling is used to interpret the formation and development of contrails. These simulations are computationally intensive and rely on high-performance computing solutions, and the contrail structures are not well defined. We propose a visual computing system to assist in defining contrails and their  characteristics, as well as in the analysis of parameters for computer-generated aircraft engine simulations. The back-end of our system leverages a contrail-formation criterion and clustering methods to detect contrails' shape and evolution and identify similar simulation runs. The front-end system helps analyze contrails and their parameters across multiple simulation runs. The evaluation with domain experts shows this approach successfully aids in contrail data investigation.
} 


\CCScatlist{ 
 \CCScat{K.6.1}{Management of Computing and Information Systems}%
{Project and People Management}{Life Cycle};
 \CCScat{K.7.m}{The Computing Profession}{Miscellaneous}{Ethics}
}

\teaser{
  \centering
  \includegraphics[width=\linewidth]{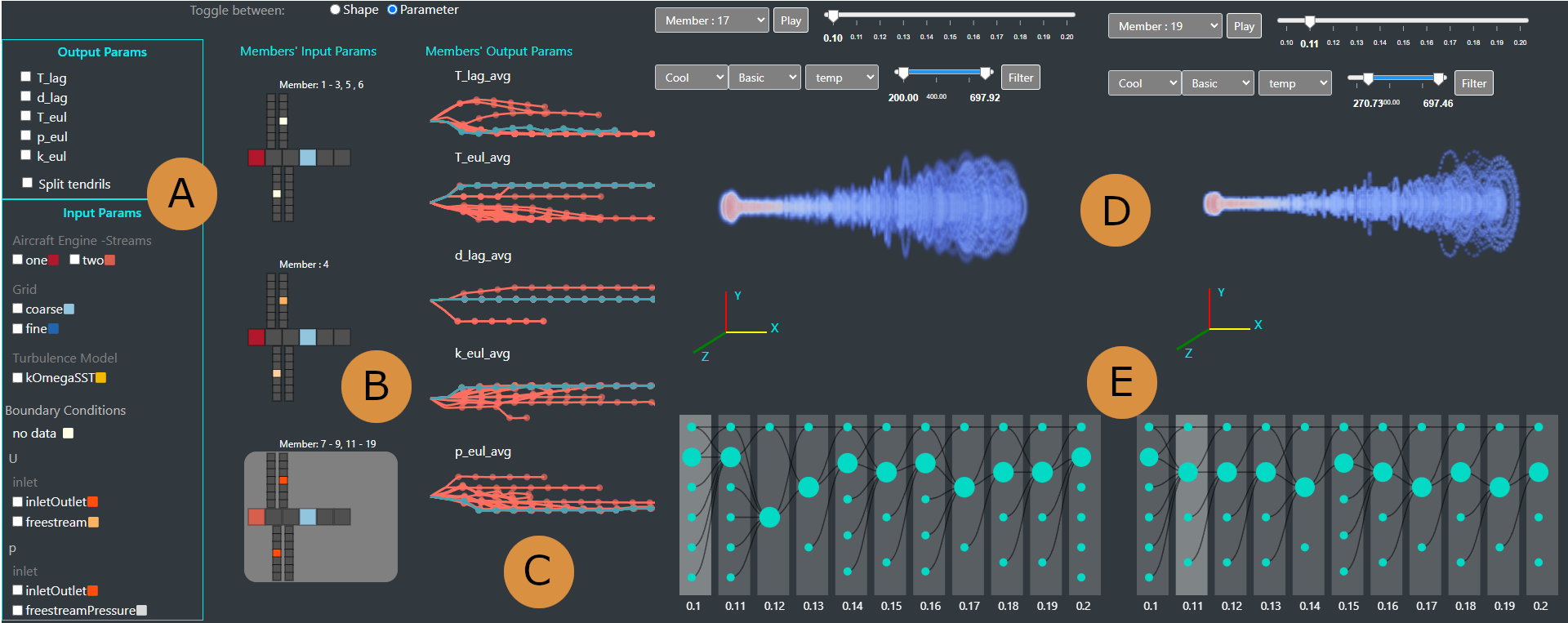}
  \caption{\textcolor{black}{Visual analysis of contrails using multiple coordinated views. (A) Filters for I/O/model parameters and color legend. (B) Colored-tile glyphs representing groups of same-parameter ensemble members; the bottom run is currently selected. A large number of runs appear to share the same input parameters. (C) Filament plots grouping output parameters of all ensemble runs; the current run is highlighted and appears significantly different. (D) 3D plumes of two similar runs, showing cooling temperatures further down the jet plume, along with animated particle trajectories. (E) Contrail evolution panel displaying the progression of contrail structures, showing a general trend where smaller groups merge into larger groups.} }
  \label{fig:teaser}
}

\vgtcinsertpkg

\begin{document}


\firstsection{Introduction}

\maketitle

Aircraft engines emit hot gases and particulates such as carbon dioxide, water vapor, hydrocarbon, and soot particles. Under sufficiently low ambient temperature, the soot particulates form visible white lines in the sky by condensing water vapor. These visible white lines become additional ice clouds in the form of condensation trails called contrails. Under the right conditions, contrails can spread up to several square kilometers and become indistinguishable from natural clouds. Like regular cirrus clouds, contrail cirrus clouds have two competing effects on climate. They shade the Earth by reflecting incoming sunlight into space. At the same time, they trap heat radiating from the Earth’s surface, particularly at night, causing warming of the air below. Contrails heighten the effect of global warming, accounting for more than half (57\%) of the entire climate impact of aviation~\cite{bbc}. With the growing air traffic leading to an increase in aircraft emissions of contrails over the last two decades, scientists are trying to find how and to which extent engine architectures, different fuels, and atmospheric conditions contribute to climate change~\cite{contrails2016paoli}.

\textcolor{black}{Computational Fluid Dynamics (CFD) experts analyze contrail formation using computational aircraft engine simulation models~\cite{wake2004paoli}. These models are complex and require high-performance computing (HPC) power. Each simulation model is run several times with different input parameters and boundary conditions to generate spatio-temporal, multivariate output, often in the form of ensemble data. Because of the data complexity, domain experts leverage pre- and post-processing to balance computational and human effort. As part of post-processing, they often seek the help of data visualization to interpret the output---Paraview is used to calculate and visualize basic quantities related to, for example, particle diameters, and Python scripts are used to average and plot, for example, the ice particle radius as a function of the distance from the jet (see Supplemental Materials). These basic explorations do not support, however, the comparative analysis of ensemble members, do not capture the contrail shape or relationships among the input, output, and the model used, nor do they help define the contrail characteristics.
}

The visual analysis of contrail data needs to meet several challenges. First, due to the complexity of the problem, contrail simulations depend on an unusually large number of parameters ($>$30), and visualizing these many parameters in a meaningful way is difficult. Second, the contrail problem requires complex computational modeling, which tends to introduce modeling errors---for example, our collaborators have been experimenting and struggling with details in the computational models for several years. Thus, creating visualizations that can expose such modeling errors is necessary. Third, identifying the contrail and characterizing its structure and evolution in meaningful visual ways is difficult due to the lack of an appropriate computational approach. Fourth, detecting similarities among simulation runs likely depends on the spatial constructs produced at the output, which are not well defined. Last, the HPC-generated data is very large \textcolor{black}{($>$100\, GB, or rather GiB)}, which poses a challenge to effective visualization.

In this work, we present an interactive visual computing framework to analyze contrail ensemble data---in particular, multiple airplane engine simulation runs. The system back-end leverages clustering methods to detect the shape and evolution of contrails, and then group together similar simulation runs. The system front-end provides the means to analyze contrails and their parameters over time.

The contributions of this work are: 1) a description of the application-domain data and tasks, with an emphasis on quantifying the contrail spatial features and identifying their similarities; 2) a characterization of the properties and criterion of contrails formation, and a description of an algorithm to detect contrail shapes and their characteristics; 3) the process of blending data mining and interactive visual encodings to explore contrail trends through clustering, as well as contrail evolution using a customized tracking graph and 3D views; 4) an implementation of the resulting design in a novel visual analysis system; 5) an evaluation by domain experts, showing the effectiveness of the system.

\section{Related Work and Background}
\noindent\textbf{Contrail Background.}
 Aircraft emissions alter the chemical composition of Earth’s atmosphere by creating ice clouds known as contrails~\cite{schumann2005formation}. The transformation of contrails from initially line-shaped to indistinguishable natural clouds occurs throughout the life cycle of the contrail, which is represented by four regimes~\cite{gerz1998transport, effects2013paoli}. The first regime, the jet regime, occurs a few seconds after emission; the second one, the vortex regime, occurs minutes following the first phase; next, the dissipation regime follows minutes after the vortex regime; and finally, the diffusion regime continues a few hours after the dissipation regime~\cite{effects2013paoli}. Due to these multiple regimes, contrail modeling is challenging, and requires multiple resource-intensive HPC-generated simulations with many parameters \textcolor{black}{($>$30)}. 
 Environmental research has been carried out for the different regimes of the wake, using in situ measurements~\cite{evolution1998spinhirne} or satellite observations~\cite{regional2002meyer} to understand contrail formation and its effect on the environment. 

This work mainly looks at the jet regime.
Paoli et al.~\cite{wake2004paoli} have observed that in this regime contrails start forming at the engine's edge, at lower temperature and higher humidity. 
However, the main focus of previous research was on simulation modeling and achieving optimal simulation parameters for contrail formation. In contrast, we present a visual analytics framework to analyze the simulation input, model, and output (I/O) parameters, to compare simulation runs, and to support the in-depth analysis of contrail formation and its longitudinal progression. 

\noindent\textbf{Computational Fluid Dynamics and Natural Science Visualization.}
Data visualization plays an essential role in the scientific studies of natural phenomena, from fluids to dark-matter~\cite{techniques2002adabala,virtual1991bryson, marai2019immersive, texturing2002ebert,particle1983reeves, hanula2015cavern, hanula2019darksky}. Analyzing spatio-temporal relationships in these data often requires extracting features and exploring attributes~\cite{aigner2007visualizing, visual2006mehta}. Similarly, in this work, we detect and extract contrail-related spatio-temporal features.

Multiple coordinated views are a common technique used in computational fluid dynamics (CFD) to analyze regions of interest while reducing visual occlusion~\cite{interactive2003bonneau}. Demir et al.~\cite{multicharts2014demir} distinguish spatial locations and variations of the flow data by analyzing the statistical properties of 3D ensemble fields. However, they only considered the whole data distribution. In contrast, we consider both individual and overview levels of the data. 

In natural science visualization, volume-renderings\textcolor{black}{~\cite{levoy1988display}} are incorporated to reveal important features of 3D fields. Liu et al.~\cite{gaussian2012liu} created a single volume by modeling the ensemble members as a Gaussian Mixture Model at each grid point. Lukasczyk et al.~\cite{viscous2017lukasczyk} applied Gaussian filtering and direct volume rendering to identify viscous fingers from salt concentration.  In this work, we follow a similar approach to obtain a 3D representation, and we use direct volume rendering to identify contrail-related spatio-temporal features.

\noindent\textbf{Ensemble Visualization.}
Ensemble data is a collection of outputs generated from different executions of the same simulation models with slightly varying parameters~\cite{earth2002hibbard}, or executions of different simulation models~\cite{ensemblevis2009potter, weather2005gneiting, reconciliation2014poco}. This data is usually generated to model initial boundary conditions~\cite{ensemblevis2009potter, parametric2014yan}, investigate parameters~\cite{parametric2014yan, multiresolution2017wang, timevarying2017biswas, review1994hamby}, analyze uncertainty~\cite{timevarying2017biswas, cone2022hurricane, noodles2010sanyal} or compare different ensemble models~\cite{ensemblevis2009potter, reconciliation2014poco}. Due to the advancement of computational power and data acquisition tools, ensemble data is generated at an unprecedented rate throughout varied disciplines~\cite{earth2002hibbard, analysis2011thompson}. Yet, its complex nature makes it difficult to analyze~\cite{survey2019wang} and visualize~\cite{exploring2009jianu, from2012potter}. 
 

Often statistical summaries such as mean, variance~\cite{curve2014mirzargar, contour2013whitaker}, modeling probability distributions~\cite{modality2016bensema, marchingCubes2011pothkow}, and clustering methods~\cite{visual2016ferstl, streamline2016ferstl} are used to reduce the complexity of the ensemble data. 
Summary-based visualization techniques such as summary statistics~\cite{cinema2016oleary}, probabilistic features~\cite{probabilistic2012petz, global2012pfaffelmoser}, color maps, contours, animation~\cite{variability2013pfaffelmoser, visualization2011coninx}, contour boxplots~\cite{contour2013whitaker}, curve box plots~\cite{curve2014mirzargar}, spaghetti plots~\cite{noodles2010sanyal}, and glyph-based visualization~\cite{flowradar2011hlawatsch, variability2013pfaffelmoser}  are used to display the overview and find relationships between ensemble members. Nonetheless,  these techniques do not work well with large data, with the analysis of many parameters, or with a detailed distribution of ensemble members. In contrast, we incorporate customized encodings to show the large ensemble data, emphasizing parameters and individual ensemble members.
 
Temporal trend analysis is an important task in ensemble visualization. Many techniques rely on either juxtaposing multiple views ~\cite{multiresolution2017wang} or superimposing plots at different time steps ~\cite{time2017ferstl} to explore temporal trends of ensemble data. However, these methods have scalability problems and can result in visual cluttering. Time-series plots~\cite{ovis2014hollt, visual2015bock}, uncertainty cones~\cite{cone2022hurricane}, and curve boxplots~\cite{curve2014mirzargar} have also been introduced. Yet, these visualizations \textcolor{black}{do not} provide details of the members at specific time points. On the contrary, our analysis focuses on the overview of the  simulation members at a given time.

Coordinated multiple-views have been adopted to analyze both input parameters and output ensemble data. These aggregated views may include multi-chart visualization~\cite{multicharts2014demir, ovis2014hollt, makingSense2020dahshan}, colored overlays~\cite{multidimensional2012biswas}, series of parallel coordinates plots~\cite{association2016liu}, or various types of tracking graphs~\cite{analyzing2010bremer, viscous2017lukasczyk, interactive2012wathsala}. Luciani et al.~\cite{details2019luciani} used multiple-linked views to explore multi-run ensemble simulation and facilitated the understanding of ensemble characteristics, but did not consider the correlation between I/O parameters or the direct comparison between ensemble members. 
Similarly, our work builds on ensemble data emphasizing the run characteristics; however, our focus is on the relation between I/O parameters and simulation output.

\begin{figure}[t!]
\includegraphics[width=\linewidth]{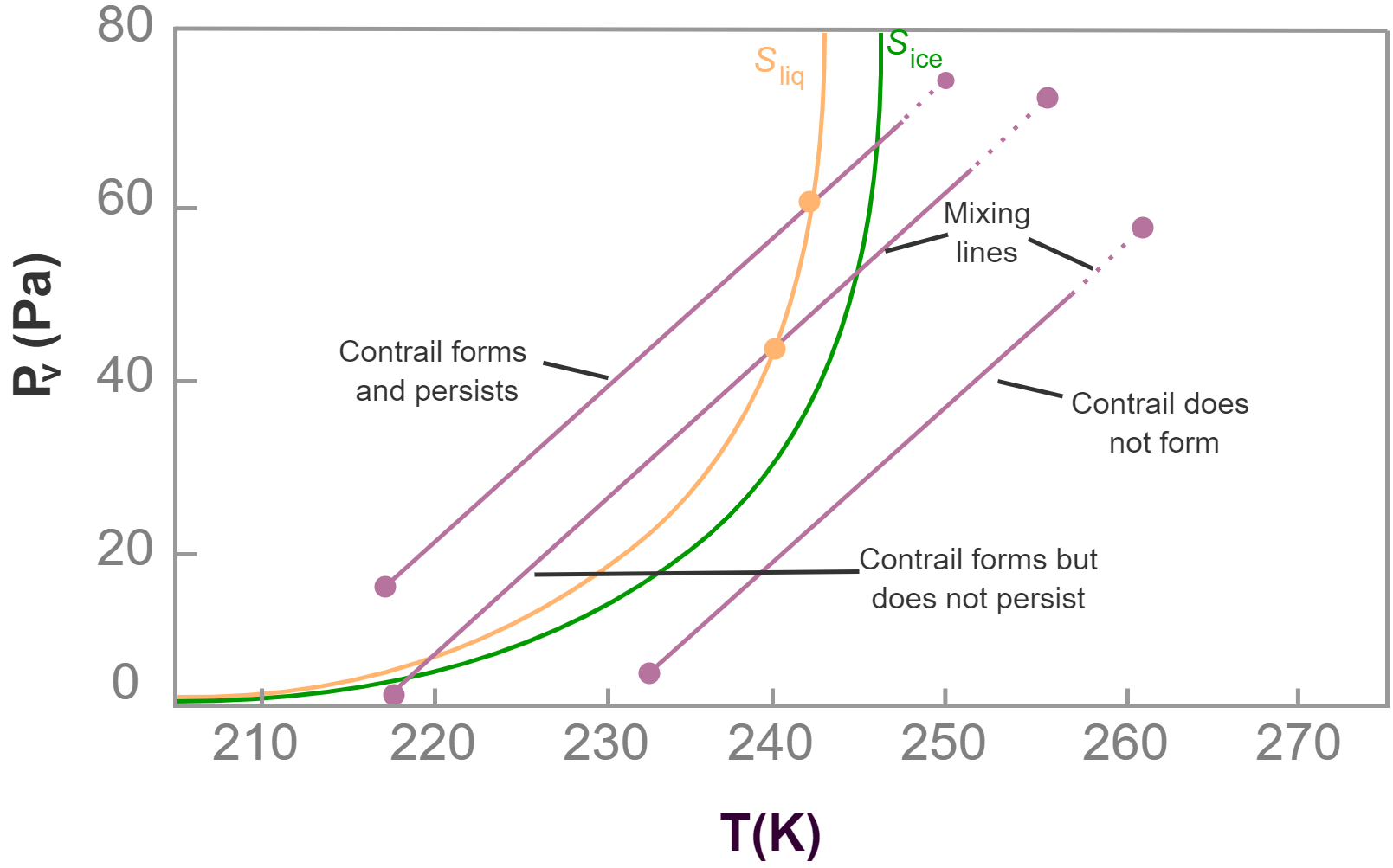}
\caption{Three scenarios for contrails formation on the temperature ($T$) - water vapor pressure  ($P_v$) diagram. Saturation curves with respect to water $S_{liq}$ and ice $S_{ice}$ are represented in orange and green colors. Each particle follows the mixing line paths from exhaust (top-left) to ambient (bottom-right) conditions. The Mixing lines show the conditions for contrail formation and their persistence in the environment.}
 \label{fig:contrail_formation}
\end{figure}

\noindent\textbf{Clustering, Distance Measures.}
In ensemble visualization, clustering methods are used to group members to reveal meaningful patterns in the data. K-means~\cite{multicharts2014demir, ensembleGraph2016shu}, hierarchical clustering~\cite{multiresolution2017wang}, and DBSCAN~\cite{comparative2016liu} are the most common clustering algorithms.
Clustering methods can use different common distance measures such as the Euclidean~\cite{visual2016ferstl, time2017ferstl, interactive2015splechtna, visualizing2018kumpf, visual2011gleicher}, the  Manhattan~\cite{visualization2004bordoloi}, the  Mahalanobis distance~\cite{mahalanobis1999mclachlan, jarema2016comparative} or application- or data-specific measures that use scalar values as the feature vectors~\cite{visualizing2018kumpf}, or sums of squared intensity differences~\cite{result2010bruckner}, etc. In this work, we incorporate traditional distance measures and domain-specific characteristics to achieve relevant clustering results for our domain problem; details of our approach are discussed in Section~\ref{subsec: computationalBackend}.

\section{System Design and Overview}

The present project was designed and developed through an interdisciplinary collaboration between two research groups over two years. Due to the 2020 pandemic, the collaboration was remote. Our team consisted of a mechanical engineering research group composed of a senior domain expert and a graduate student, and a visual computing team composed of a senior visualization expert and two graduate students. The team met on a weekly basis to discuss simulation generation, data processing, and the system’s design. \textcolor{black}{Feedback from the collaborators was incorporated into the design}.

This project used an Activity-Centered-Design (ACD) approach~\cite{marai2017activity} to design the visual computing framework. The ACD paradigm is an extension of the Human-Centered-Design paradigm, focusing on user activities and workflows. Using this paradigm, the team met over multiple sessions to determine functional specifications, prototype, evaluate encodings and interfaces, and decide upon necessary changes.

\textcolor{black}{Our computational back-end was built using Python with pandas, scikit-learn, NumPy, and  Jupyter.} The front-end was developed using JavaScript with D3.js~\cite{bostock2011d3} and WebGL.

\subsection{Activity Analysis}
\textcolor{black}{Our collaborators had extensive experience in  ParaView~\cite{paraview2015ayachit} with OpenFOAM~\cite{jasak2007openfoam} in post-processing, although they could not compare or validate multiple simulations simultaneously, nor characterize the contrail. To visualize a contrail in ParaView, they would color-code the particles by diameter size and consider any diameter larger than the initial soot diameter as an indicator of ice or contrail formation. Additionally, they manually checked the input parameters and boundary conditions to validate the simulation data, and used Python to plot basic output quantities (see Supplemental Materials). Investigating the contrail formation, defining their characteristics and their evolution, and identifying similar members was impossible due to a lack of an appropriate computational approach. Furthermore, they wished to compare multiple outputs and input/model conditions, for which capabilities were limited, even with the use of scripts. They were also concerned whether the model behaved appropriately, and they were further concerned with data quality. } 

After multiple meetings with the collaborators, we summarize the functional requirements for the project as activities below:

\begin{itemize}
\itemsep0em \vspace{-0.8em}

\item \vspace{-0.3em} A1. Explore  multiple simulation runs and determine whether the model behaves as expected
\item \vspace{-0.3em} A2. Summarize a set of simulation runs
\item \vspace{-0.3em} A3. Derive a contrail formation criterion
\item \vspace{-0.3em} A4. Characterize the contrail structures
\item \vspace{-0.3em} A5. Analyze the contrail structures temporally
\item \vspace{-0.3em} A6. Analyze simulation runs based on contrail characteristics
\item \vspace{-0.3em} A7. Identify run similarities based on input, model, and output parameters
\end{itemize}
\vspace{-0.8em}
 
The non-functional requirements included a request for an easy-to-access visual framework that efficiently handles large ensemble data, \textcolor{black} {data quality awareness, and the ability to visually handle the variability of I/O parameters across simulations.} 

\subsection{Data}
Generating the data for an ensemble member can take up to several days and requires HPC power to extract relevant information. The data are large \textcolor{black}{($>$100\, GiB), and are generated from multiple computer-generated aircraft engine simulations. Each simulation is run with different user-defined categorical input and model parameters (e.g., aircraft engine type, grid resolution, geometry, scope, etc.), and with different boundary conditions. This process generates multiple output ensemble members, also referred to as simulation runs. Each run contains details about particle trajectories, as well as numerical properties, which are used to analyze contrail formation for each engine type. Each output ensemble consists of multiple time steps (ranging from 10 to 15), and each time step features particle numerical attributes such as position, temperature, diameter, ice label, and pressure. The data consisted of a total of 29 simulation runs, where 19 of the runs had multiple time steps, and 10 of the runs yielded the final full-grown contrail structure.}

\subsection{Contrail Formation Criterion} 
We considered the physics of the problem to describe the process of contrail formation, and to define its structure (A3). In high relative humidity, for example in the tropics, and at high altitudes, contrails are formed in the jet plume when moist and unsaturated hot-exhaust gasses mix with the cold ambient air~\cite{formation1953appleman}. This mixing process is illustrated in the water-vapor partial pressure-temperature plot shown in Fig.~\ref{fig:contrail_formation}. Assuming that the vapor and heat diffuse at the same rate and the flow is adiabatic (i.e., a process without transfer of heat to or from a system), the mixing can be represented as a straight line called a mixing line. The saturation curves for liquid water $S_{liq}$ and ice $S_{ice}$ can then be derived from the Clausius-Clapeyron equilibrium equations for a perfect gas as follows:

\begin{equation}
   \textcolor{black}{ \frac{d \, \mathrm{ln} \, S_{liq}}{dT} = \frac{E_{liq}(T)}{RT^2}}
\end{equation}

\begin{equation}
 \textcolor{black}{\frac{d \, \mathrm{ln} \, S_{ice}}{dT} = \frac{E_{ice}(T)}{RT^2}} 
\end{equation}

\noindent where $E_{liq}$ is the latent heat of evaporation, $E_{ice}$ is the latent heat of sublimation (i.e., the transition of a substance from its solid state to directly to its gas state), $T$ is the temperature, $R$ is the molar gas constant and \textcolor{black}{$R\,= \, 8.31 \, \mathrm{J \, mol^{-1} \, K^{-1}}$}.

When a fluid element traverses through the mixing line, it first reaches ice saturation, then liquid saturation  (Fig.~\ref{fig:contrail_formation}). If the mixing line does not cross the liquid saturation curve, the contrail will not form. On the other hand, if the mixing line crosses the liquid saturation curve, the contrail will form and persist. However, after crossing the liquid saturation, if the mixing line crosses the liquid saturation again, the contrail will form but will not persist.

\begin{figure*}[t!] 
\centering
\includegraphics[width=1\linewidth]{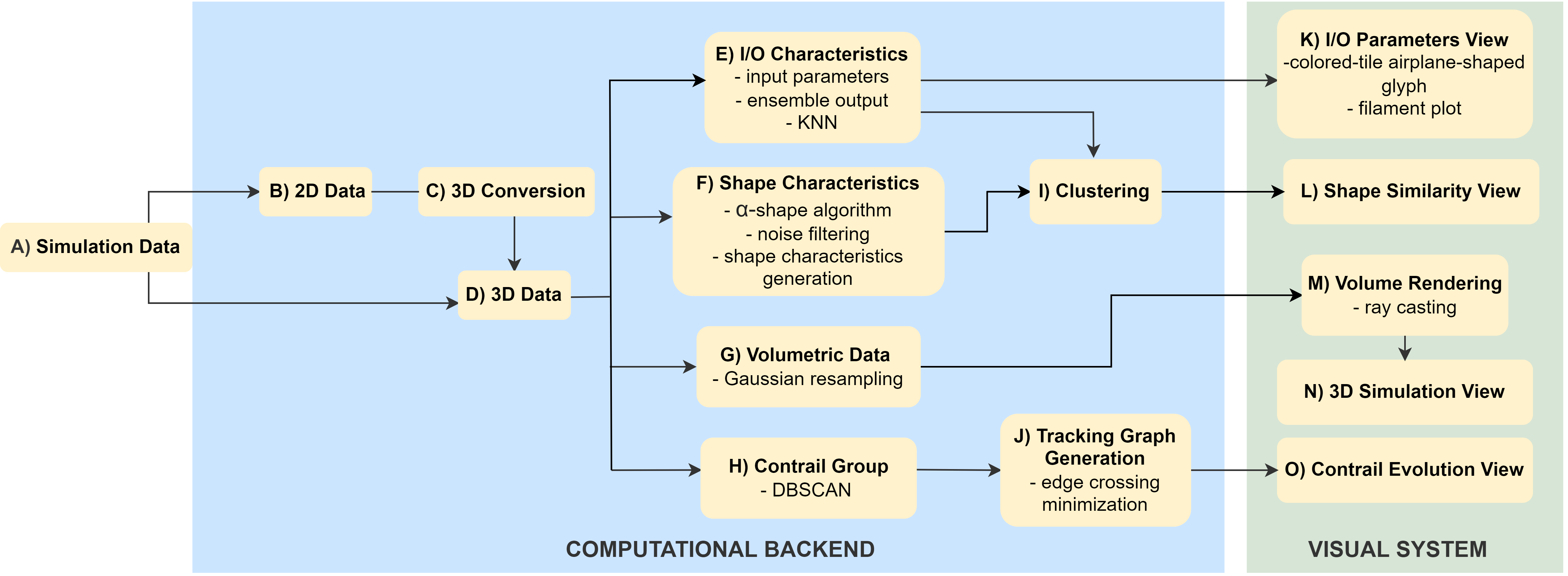}
\caption{Solution overview. The system has three main components: simulation data, computational back-end, and visual front-end. Simulation data includes the simulation input and model parameters and outputs. \textcolor{black}{The simulation data (step A) goes through several offline transformations and computations in the computational back-end (steps B-J), and then it is used in the four main views of the visual front-end (steps K-O)}.
}
\label{fig:data_workflow}
\end{figure*}

\subsection{Computational Back-end}
\label{subsec: computationalBackend}
The complex simulation data goes through several transformation and computation steps in the back-end \textcolor{black}{(Fig.~\ref{fig:data_workflow}, steps B-J)} before it can be used in the visual computing framework (Fig.~\ref{fig:data_workflow}, steps K-O). First, the data (Fig.~\ref{fig:data_workflow}, step A) is pre-processed (Fig.~\ref{fig:data_workflow}, steps B-D), and then several algorithms are applied (Fig.~\ref{fig:data_workflow}, steps E-J) to extract relevant information for the visual front-end. \textcolor{black}{Pre-processing can take up to 30 min on an 8GB Nvidia GeForce RTX 2070 GPU, and Intel 3.6GHz CPU machine, depending on the size of the simulation runs.} The data workflow in the computational back-end is presented below.

\subsubsection{Data Conversion.} \textcolor{black}{The simulations can be run on 2D grid structures (Fig.~\ref{fig:data_workflow}, step B) or 3D grid structures (e.g., cylindrical grid~\cite{cylindrical2015wang}) (Fig.~\ref{fig:data_workflow}, step D). In this work, we deal with both types of grid structures, 2D and 3D, where the simulation captures particle trajectories up to a few seconds after they exit the aircraft engine jet. Because the contrail problem is symmetric, whenever 2D data was provided,} we converted the particle data to 3D (Fig.~\ref{fig:data_workflow}, step C) by rotating the axis aligned with the jet axis; in our case, the X-axis. 

\textcolor{black}{The 3D reconstruction of the jet plume based on 2D data is a legitimate post-processing methodology to study the contrail dynamics. The flow-field is statistically two-dimensional and axis-symmetric, which means that, when averaged over many (ideally infinite) realizations, it will only depend on the axial and radial coordinates. Even if a single realization (an instantaneous ``picture'') of the contrail would still be three-dimensional, the Lagrangian particle tracking implemented in the solver is able to capture these effects in a 2D setting. This is due to the random motion of particles seeded in the flow, which span regions with different levels of temperature and vapor concentration. Examining the result in 3D is necessary even with 2D grids because the output is judged in relation to the physical phenomenon observed in nature.}

\subsubsection{Ensemble Member Similarity Measurement}
\label{subsubsec: member_similarity}
\textcolor{black}{To support activities A6 and A7, we implemented two approaches to identify similar ensemble members, based on our collaborators' requirements. The first approach identifies similar members according to the I/O and model parameters (A7)  (Fig.~\ref{fig:data_workflow}, step E). The second approach finds similar members based on the shape of the contrails (A6) (Fig.~\ref{fig:data_workflow}, step F).}

\noindent\textbf{Contrail Attribute Computation.}
Given the contrail formation criterion, for each time step, we extracted the following contrail characteristics: mean temperature, number of ice particles, the total mass, and the length of the ice (Fig.~\ref{fig:data_workflow}, step E). \textcolor{black}{Our collaborators and we arrived at this set of characteristics through repeated exploration of the data.} These characteristics enable us to analyze the correlations and similarities between input and model parameters and output. We determined for each time point the mean temperature and the number of particles that have turned into ice, and the total mass of the ice particles via the following formula:

\begin{equation}
m_{total} = \sum_{p=1}^{N} \frac{1}{6} \, \pi \, d_p^3 \, \alpha
\end{equation}

\noindent where $m_{total}$ is the total mass of the ice particles for a specific time step, $p = 1,2,..N$ is the number of particles, $d_p$ is the diameter of \textcolor{black}{$p\text{-}th$} particle, and \textcolor{black}{$\alpha = 917 \, \mathrm{kg/m^3} $} is the ice density. The mass was computed for each ice particle $p$, and summation was used to get the total mass of all ice particles for a given time step.

Next, the total length of the contrail was computed to identify how far the ice particles were spread out in the simulation environment. For a specific time step of a given simulation, if the X-coordinates of the particles were the same \textcolor{black}{as in the previous step}, we calculated the pairwise distance of all particles. Otherwise, we used the convex hull of ice particles (i.e., the smallest convex set that contains all of the points in the set~\cite{quickhull1996Barber}) to calculate the pairwise distance between the points on the hull.
Finally, we considered the two furthest particles and calculated their Euclidean distance as the length of the ice/contrail structure. 

\begin{figure*}[ht] 
\centering
\includegraphics[width=1\linewidth]{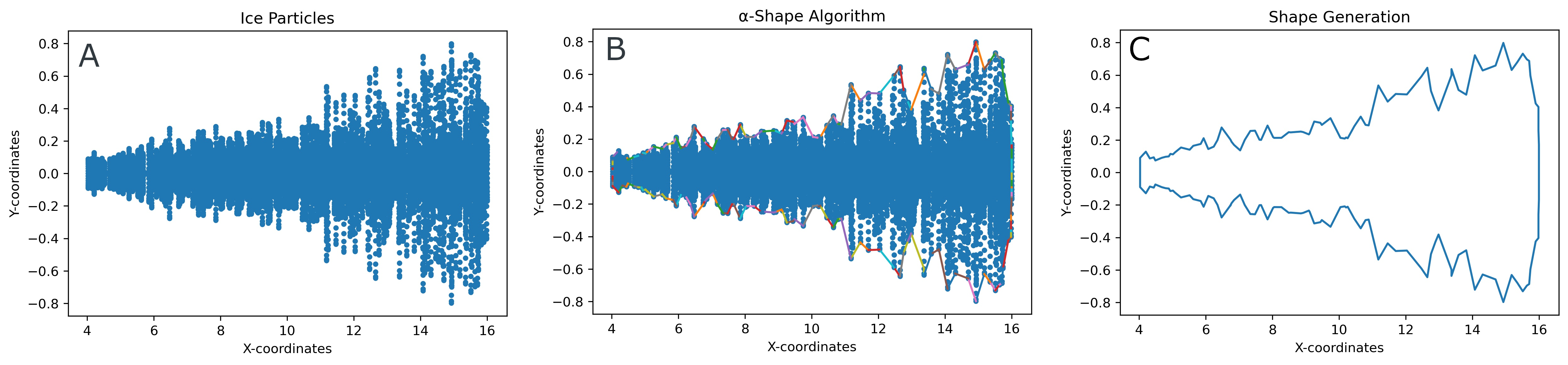}
\caption{Shape Detection Algorithm: (A) Positions of 2D ice particles from one time point of a simulation. (B) Result of the $\alpha$-Shape Algorithm, constructing points at the border of the simulation spatial output. (C) Extracted boundary points are ordered chronologically to generate the contrail shape.}
\label{fig:contrail_shape}
\end{figure*}

\noindent\textbf{Similar Members Based on I/O and model Parameters.} Analyzing the I/O and model parameters of similar ensemble members can help domain experts  achieve ideal parameters for simulation models (A7). \textcolor{black}{Our collaborators were particularly interested in this approach, as it could save them considerable time and resources when running the HPC simulations.} We used the k-Nearest Neighbor (KNN) clustering algorithm to identify related members across different simulation runs based on their I/O and model parameters (Fig.~\ref{fig:data_workflow}, step E). The input and model data consists of categorical attributes such as aircraft engine streams, scope, grid, and solution, whereas the generated contrail output consists of numerical attributes such as temperature, number of ice particles, and the total mass and length of the ice structure. Hence, selecting the right distance metric is essential to achieve accurate results. To solve this, we used the Gower distance, a measure to find the similarity between datasets consisting of mixed type attributes~\cite{general1971gower}. The Gower distance $GD_{xy}$ of two ensemble members $x$ and $y$ is calculated as the average of partial closeness across $n$ attributes: 

\begin{equation}
GD_{xy} = \frac{1}{n} \sum_{i = 1}^{n} P_{xyi}
\end{equation}

\noindent where $P_{xyi}$ denotes the partial similarities of attribute $i$. The distance between two members is the average of all attribute-specific distances; ranges between $0$ and $1$. For a numerical attribute $i$, the partial similarity between two members $x$ and $y$ is:
\begin{equation}
P_{xyi} = 1- \frac{\left| V_{xi} - V_{yi} \right|}{max(V_i) - min(V_i)}
\end{equation}
\noindent where $V_{xi}$ and $V_{yi}$ are the attribute values of member $x$, $y$ and the range of the attribute $i$ is between $max(V_i) - min(V_i)$. For a categorical attribute, the partial similarity between two members $x$ and $y$ is $1$ if both members have the same value and $0$ otherwise.

\begin{equation}
  P_{xyi} = \begin{cases}
    1, & \text{if $V_{xi} = V_{yi}$}\\
    0, & \text{if $V_{xi} \neq V_{yi}$}
  \end{cases}
\end{equation}

Our algorithm for finding similar ensemble members based on I/O and model parameters is the following: 1)  we take the I/O and model parameters as attributes for all members, throughout all simulation runs; 2) we calculate the Gower distance for each pair of members, which is then used as the distance metric for the KNN clustering method; 3) we compute the five most similar members based on the I/O and model parameters for each member.

\noindent\textbf{Contrail Shape Detection.} Determining the shape and spread of the contrail was important to \textcolor{black}{the domain experts, who wished to} understand the evolution of contrail structures and their spatial characteristics (A6). In computational geometry, many algorithms have been introduced to compute the shape of a set of points. Jarvis~\cite{onthe1973jarvis} proposed an algorithm to estimate shape as a generalization of the convex hull of a planar point set. Edelsbrunner et al.~\cite{shape1983edelsbrunner} later introduced a mathematical definition of shape and proposed an algorithm to find the shape of a planar point set called \textcolor{black}{$\alpha$-shape}. Their algorithm is based on Delaunay triangulations~\cite{two1980lee}. In this work, we incorporate Eldesbrunner’s definition and calculate the 2D contrail shape based on the \textcolor{black}{$\alpha$-shape} algorithm (Fig.~\ref{fig:data_workflow}, step F).

Our procedure is: 1) for a specific time step of a simulation run, consider the $(X,Y)$ coordinates of the ice particles (Fig.~\ref{fig:contrail_shape}.A); 2) calculate the \textcolor{black}{$\alpha$-shape} for that time step, where we only preserve the outer border and return a set of pairs representing the edges of the $\alpha$-shape (i.e., a set of points at the boundary) (Fig.~\ref{fig:contrail_shape}.B); 3) chronologically add the points to a list to get all the points contributing to the shape of the contrail (Fig.~\ref{fig:contrail_shape}.C).

\noindent\textbf{Similar Members Based on the Contrail Shape.} \textcolor{black}{We apply KNN to each member to determine the five most similar ensemble members based on their shape (A6). Initially, we used the Hausdorff distance~\cite{efficient2015taha}, which is commonly used to determine similarities between two object shapes~\cite{comparing1993huttenlocher}  (Fig.~\ref{fig:data_workflow}, step F). Yet, because this measure is susceptible to outliers and noise data, it did not yield accurate results in our case. Moreover, our collaborators wished to consider shape characteristics such as the area of the shape,  its length, height, and slope. Therefore, we later followed a different approach where we first applied noise filtering and then extracted the shape characteristics.}  

\noindent\textbf{Noise Filtering and Similar Shape Detection. } \textcolor{black}{There are cases where a few particles veer off from the contrail structure and still turn into ice. These distant particles affect the overall shape of the contrail as determined by the \textcolor{black}{$\alpha$-shape} algorithm  (Fig.~\ref{fig:contrail_shape}), and thus also the subsequent similar shape detection results. The experts agreed that such particles can be safely filtered out. To this end, we performed least squares linear regression on the 2D projected points in the structure's upper half. A distance threshold from the regression line was determined heuristically as five times the vertical standard deviation of all the ice particles in that time step. The experts agreed that particles higher than this threshold (respectively lower in the symmetric lower half) could be safely skipped. After removing this noise and consulting with the experts, we then extracted the characteristics for each shape, and, for each member, we applied KNN to identify the five most similar ensemble members.}

\begin{figure*}[t!] 
\centering
\includegraphics[width=1\linewidth]{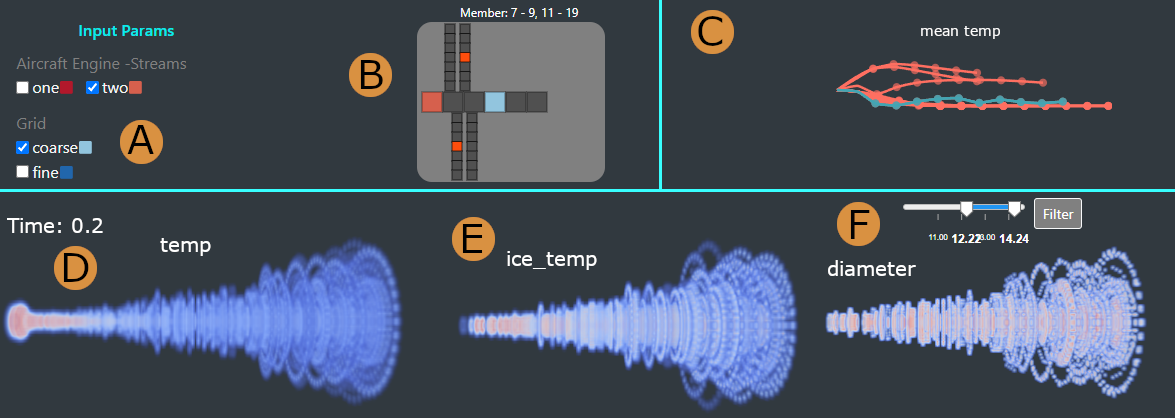}
\caption{Run analysis. (A) Filtering by two-stream aircraft engine and coarse-grid reveals (B) a large group of simulation runs with these settings (see the color legend in A and the tile glyph in B). (C) Two groups are also revealed, with low mean temperature in the filament plot. The distributions of (D) soot particles, and (E) ice particles for a specific time step can be further examined. The filtered simulations show that contrails start to form further from the jet exit. (F) Filtering the simulation based on diameter values reveals that most ice particles have large diameters. }
\label{fig:filtered_simulation}
\end{figure*}

\subsubsection{Contrail Group Detection and Tracking}
For an in-depth exploration of contrail formation, in each simulation run, we identified particles that turned into ice and the number of ice clusters present at each time-step (A4) (Fig.~\ref{fig:data_workflow}, step H). \textcolor{black}{The experts were interested in identifying large contrail structures, their characteristics, distribution, and evolution over time. }

\textcolor{black}{We applied the DBSCAN algorithm~\cite{density1996ester}  to determine the number of ice clusters present in a specific time step, as DBSCAN can discover clusters of arbitrary shapes based on the spatial density of ice particles \textcolor{black}{(see the Supplemental Materials)}. The model generated using DBSCAN is dictated by the threshold distance $eps$, which is used to determine whether two points are neighbors or not.} 
\textcolor{black}{To determine the optimal $eps$ value, we used a algorithm similar to the one by Rahmah et al.~\cite{determinaiton2016rahmah}, which calculates the Euclidean distance of each pair of particles. For each particle, we sort the closest distance to the neighbors in ascending order and then consider the distance of the k nearest neighbors (k=3). The distance values are then represented as a curve, where $eps$ corresponds to the point of maximum curvature (i.e., critical change in the curve), which is calculated using the Satopaa et al. technique~\cite{kneedle2011satopaa}.} 

In short, our proposed algorithm to identify contrail structures (Fig.~\ref{fig:data_workflow}, step H) is the following: 1) for a specific time point, we only consider the particles that have turned into ice; 2) we calculate the optimal $eps$ value to run the clustering algorithm; 3) we run the DBSCAN algorithm to find the clusters (i.e., contrail structures) for the present time step; 4) we repeat the above steps for every time step of a simulation run. 

\textcolor{black}{Next, we process the contrail group structures that were identified previously, in order to track their progress across a simulation run (see Supplemental Materials). This process determines the temporal context of the contrail evolution and captures the formation, dissipation, merging, and splitting of contrail structures. In addition, this process helps to determine the number of particles in each group and their characteristics (e.g., the length or mass of different clusters). }



\subsection{Visual Front-End}
\textcolor{black}{We used a parallel prototyping approach~\cite{dow2010parallel}, due to its proven success in making better design choices, and in stimulating more detailed and constructive feedback compared to serial prototyping. The design of our framework is based on multiple coordinated views, which support  both overview and details, and provide the ability to visually integrate multivariate spatio-temporal ensemble data (A1). The final design leverages qualitative feedback from our collaborators.} The visual front-end consists of four main views: 1) An Input and Output Parameter View (Fig.~\ref{fig:teaser}.B, C) assists in providing a summary of the simulation run parameters (A2, A7); 2) A 3D Plume Projection View (Fig.~\ref{fig:teaser}.D) supports the examination of simulation particles and contrail formation (A1, A2); 3) A Contrail Evolution View (Fig.~\ref{fig:teaser}.E) allows exploring the contrail temporal progression (A4, A5); and 4) An alternative Similar Shape Exploration View \textcolor{black}{(Fig.~\ref{fig:member_similarity}.A)} aids in the identification and analysis of similar simulation runs (A6, A7).

\subsubsection{Input and Output Parameter View} 
This view provides a guided summary of the ensemble members’ simulation parameters. The view helps identify areas of interest across multiple simulation runs. It also helps select, compare, and examine specific ensemble members (A2, A7) (Fig.~\ref{fig:teaser}.B,C). The  input, model and output parameters are displayed in two linked panels. A filtering panel allows filtering members across the interface (Fig.~\ref{fig:teaser}.A). 

Finding a suitable visualization for a large number of categorical input variables ($>$30) proved to be challenging. \textcolor{black}{After multiple attempts, we arrived at a custom colored-tile glyph (Fig.~\ref{fig:teaser}.B) that maps the parameters to what was originally an airplane-inspired shape (e.g., aircraft attributes on the fuselage, simulation boundary conditions on the first left-wing, particle attributes on the first right-wing, etc), but evolved into a general two-dimensional glyph. We designed this glyph (see Supplemental Materials), intuitive to the experts, through multiple feedback-driven prototypes that explored rich, dynamic shapes and colors.}

\textcolor{black}{Visualizing the large set of categorical values with the glyph was still a challenge, as the color-dense resulting display was overwhelming to the experts.  Fortunately, we realized that many simulation runs had the same input parameters. Our collaborators agreed that they were mainly interested in the effect of the differences in parameters between members. Therefore, instead of showing increasingly complex tile glyphs with various colors, we emphasized the \textit{differences} between parameter settings for all simulations through color. This approach provided an elegant solution to the challenge of large numbers of input parameters. Furthermore, this representation enabled us to group ensemble members with the same parameters, and show a single group representative glyph.} 

Glyph tooltips provide additional details about each attribute. \textcolor{black}{To reduce cognitive load, a color legend for the glyph is provided within the input parameter panel. The panel consists of the attribute names, their values, and the glyph color mapping (Fig.~\ref{fig:teaser}.A).}

The output simulation parameters are sequential event-based attributes. Mean values for each time step were computed for these attributes.\textcolor{black}{We group the ensemble members per parameter, for easier progression interpretability. The design process explored a wide range of possible temporal encodings, many of which showed scalability problems. After several sessions, the design process focused on a promising encoding called a filament plot~\cite{floricel2021thalis}. Filament plots emanate from a common root, then proceed in a left-to-right direction aligned with the time sequence. We used a plot for each attribute, where each filament represents the entire observation period for an ensemble member's output attribute, with dots along the filament to indicate time stamps. The currently selected run is highlighted with color. To account for inter-member attribute variability, the curvature degree for the filament at each time step encodes the relative change from the previous attribute value, where upward rotation indicates value increases and vice versa for the downward rotation.}

\begin{figure*} 
\centering
\includegraphics[width=1\linewidth]{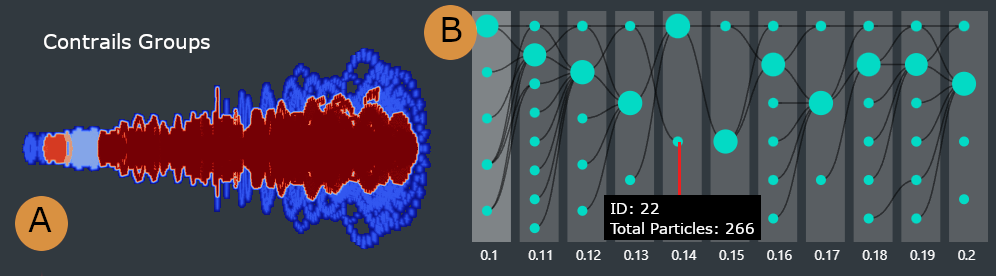}
\caption{Contrail structure groups and evolution over time. (A) Different colors showing the contrail structures of the 0.1 time step of a simulation run. The largest contrail structure is at the center and further away from the jet exit. (B) Contrail structure evolution over time showing the merging/splitting of groups. The highlighted group does not move to the next time step, indicating the group leaves the simulation environment.}
\label{fig:contrail_groups}
\end{figure*}

\subsubsection{3D Plume Projection View}
To support the evolution and comparison of two simulations, the 3D plume projection view (Fig.~\ref{fig:teaser}.D) displays the particle distribution over time, color-coded based on a set of pre-selected attributes such as temperature or diameter (A1, A2). The plume denotes the aircraft jet engine exhaust.

We applied direct volume rendering (Fig.~\ref{fig:data_workflow}, step G) to analyze the 3D simulation of ensemble members over time, emphasizing the spatial features, formation, and evolution of the contrail structures (A2) that are released through the plume. As each simulation run can contain up to millions of particles, the slow rendering time can affect the system's overall efficiency,  \textcolor{black}{which was a concern for the domain experts}. Therefore, we incorporated volume rendering  to efficiently show the detailed distribution of the ensemble data without compromising important information. Our method works well as the size of the volumetric data does not increase with the number of particles. Based on the contrail criterion \textcolor{black}{and feedback from the domain experts}, we used the temperature, the diameter, the ice label, and the group of contrail particles to calculate the density distributions of the particle data. After extracting the volumetric data, we applied a direct volume rendering technique---namely, ray casting---using WebGL, which provided better image quality than other methods. 

Additionally, \textcolor{black}{as required by the domain experts,} animations of all simulation runs show the progression of contrails throughout the whole simulation, and filtering options are available based on the particles’ physical properties (Fig.~\ref{fig:filtered_simulation}.F). Different shader options are provided for better readability of the simulation results based on their attributes (e.g., particle temperature). \textcolor{black}{We used the MIP shader to identify the distribution and overall shape of the contrail group structures (Fig.~\ref{fig:contrail_groups}.A). Additional shaders can be selected to render the clusters of particles, and the filtering slider can be used to filter the particles based on their cluster group.} A time slider allows exploring specific time steps.

\subsubsection{Contrail Evolution View}
The contrail evolution view (Fig.~\ref{fig:teaser}.E) displays the progression of contrail structures during  a simulation run through a customized node-link diagram (A4, A5). \textcolor{black}{In particular, the domain experts wished to understand contrail evolution, and whether and how contrail structures merge.} We first tried a node-link encoding with a forced layout because it correctly captured the contrail grouping. However, it did not preserve the temporal aspect of the data, so we customized it to follow a left-to-right direction to better emphasize the temporal aspect of the structures’ evolution. Each column corresponds to a time step during a simulation, while each circle node represents a contrail group structure. The radius of each node is scaled by the number of particles in that group. The group of contrails may merge or split according to the longitudinal evolution of the contrails (Fig.~\ref{fig:contrail_groups}.B), which is emphasized by the diagram’s links. We minimize edge-crossing by implementing a simple algorithm that starts from the most recent timestep and traverses backward, positioning each child  group near its parent. If a child has multiple parents, the algorithm positions the child near its most recent parent. Finally, hovering over a child shows details-on-demand information about the group id, mean temperature, number of ice particles, mass, and length.

\subsubsection{Similar Shape Exploration View}
\begin{figure*} 
\centering
\includegraphics[width=1\linewidth]{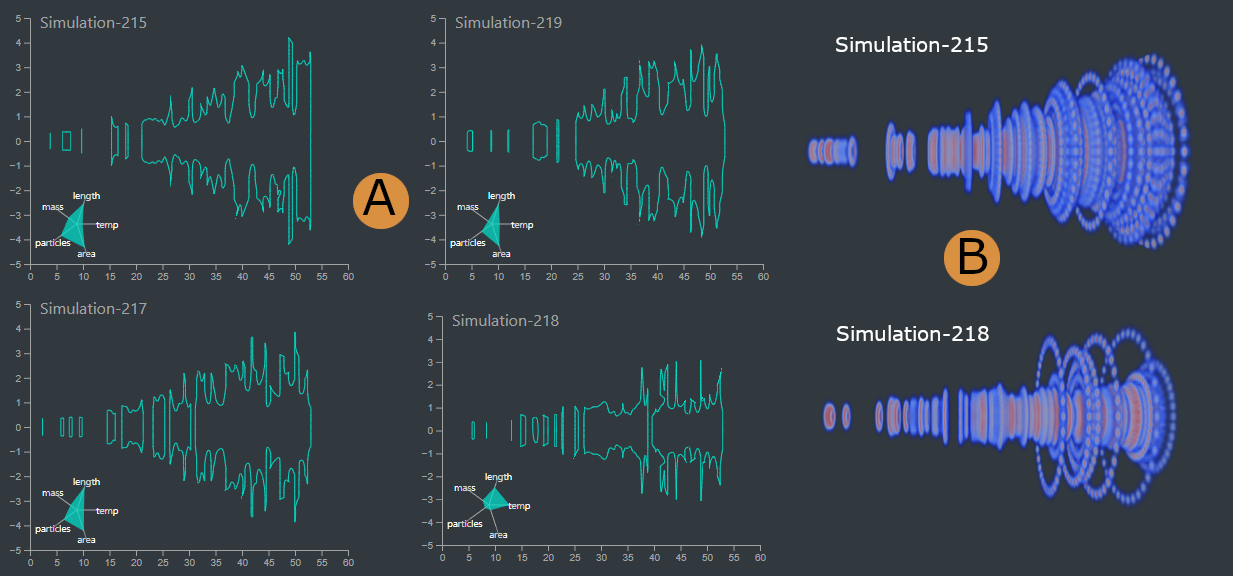}
\caption{Contrail Shape Similarity Exploration. (A) Shape of the contrail generated by Simulation 215, and the three most similar members based on shape characteristics. Kiviat diagrams show the contrail characteristics of the simulation. It can be noticed that member 218 has a narrower shape and different characteristics than the other three simulation runs. (B) 3D shape of a wide-shaped simulation (top) and a narrow-shaped simulation (bottom).
}
\label{fig:member_similarity}
\end{figure*}

An alternative panel to the input and output parameters panel (Fig.~\ref{fig:teaser}.A-C), this view aims to identify similar ensemble members based on shape characteristics or I/O parameters (A6, A7) (Section~\ref{subsubsec: member_similarity}) \textcolor{black}{(Fig.~\ref{fig:member_similarity}.A)}. For a selected member, this view facilitates the exploration of the five most similar members and their shapes based on user-defined similarity measures. Additionally, Kiviat diagrams display details about each member’s contrail attributes (e.g., mean temperature, total area, total length, total mass, and total particles). Due to their closed polygon shape, a preattentive feature, Kiviats are particularly effective in small multiple form~\cite{precision2019marai, thomas2017echo}, emphasizing similar ensemble members or outliers. Hovering over the Kiviat shows related information about the contrail characteristics of a particular member.

\section{Evaluation}
We evaluated our solution through a combination of multiple demonstrations and case studies involving two domain experts who are also co-authors of this paper. Our evaluators were not involved in the development of the visual framework at all stages, but provided qualitative feedback regularly, through weekly online meetings. Apart from the regular design feedback sessions, we completed two case studies remotely, using screen sharing and note-taking along with the think-aloud method. The exploration of the interface was directed by the domain experts, and the first author following their instructions. We analyzed 19 simulation runs during the first case study, and another 10 simulation runs for the second case study.

\subsection{Case study 1: HPC Data Quality Validation and Contrail Evolution Analysis}
\textcolor{black}{This case study focused on exploring the correctness of the numerical model and simulation, and visually validating the contrail detection approach (A1, A2, A3). To achieve these goals, our collaborators focused on examining the contrail evolution, checking the results against their practical experience, and then determining commonalities among multiple simulation runs.} 

Domain experts associate contrail formation with particle temperature; hence the exploration started by selecting one member in the 3D plume view (Fig.~\ref{fig:filtered_simulation}.D), to visually check the contrail structures as a function of temperature (A1). Before inspecting the simulation step by step, the experts checked the simulation progression as an animation. They observed that particles near the exit at the beginning of the simulation had a higher temperature, as expected, because the jet exit is very hot. Moreover, they confirmed that the particle temperature started to cool off as they traveled further from the jet exit. The animation further revealed that the mean temperature decreased over time. Next, the experts investigated the position of the contrails and the relative temperature at the final time step of the simulation (Fig.~\ref{fig:filtered_simulation}.D) by filtering the particles that turn into ice (Fig.~\ref{fig:filtered_simulation}.E). They noted that, in practice, contrails start to form at a short distance from the aircraft jet exit, where particles have cooled enough to form ice when coming in contact with water vapor. Our evaluators were pleased to notice that the detected contrail structures also started at a short distance from the jet engine. This observation was backed up by the 3D views of the other simulation runs (Fig.~\ref{fig:teaser}.D, left). Another finding was the formation of several small contrail structures close to the jet engine exit, followed by the main plume (A3). \textcolor{black}{The experts further noted that the Lagrangian model used in the simulation  determines regions of highly fluctuating supersaturation, which manifest as gaps in Figs.~\ref{fig:filtered_simulation}--\ref{fig:member_similarity}.}

Encouraged by this visual confirmation, the group selected another simulation run in the second 3D view for further validation and comparison of particles' properties with the first simulation run (Fig.~\ref{fig:teaser}.D, right) (A2). The second simulation had the same overall characteristics. When they analyzed the particle distribution based on diameter values, most particles showed larger diameters than expected (Fig.~\ref{fig:filtered_simulation}.F). From this observation, the senior domain expert noted that particles should have much smaller diameters as they turned into ice and showed his concern about the calculation for that particular simulation. Going through the simulation parameters and conditions later offline, they confirmed that the simulation used the wrong initial conditions and, thus, produced incorrect results. The domain expert mentioned that they often use ParaView~\cite{paraview2015ayachit} to manually validate the output, which can be exhaustive and prone to error. Instead, our 3D view facilitated an easy exploration and validation of the contrail structures in multiple aircraft engine simulations.

In a follow-up meeting, the domain experts wanted to evaluate parameter differences between simulation runs. They focused on the colored-tile glyphs (Fig.~\ref{fig:teaser}.B), where similar simulations were grouped based on their input and boundary conditions (A7). The evaluators immediately remarked on the large number of simulations that shared the same parameters (Fig.~\ref{fig:filtered_simulation}.B). They noted that this was unsurprising, as many simulations were run using two-stream aircraft engines and a coarse grid. Next, they observed changes in the simulation outputs over time; and moved their focus to the filament plots (Fig.~\ref{fig:teaser}.C). Specifically, they wished to understand the temperature patterns because the formation of contrails depends on whether the temperature is low enough. Filtering by low mean temperature (Fig.~\ref{fig:filtered_simulation}.C), the experts found two groups of simulation runs, with one having lower temperatures throughout the whole simulation (A7).

In the end, the group selected a member with a low mean temperature in the first 3D view (Fig.~\ref{fig:teaser}.D). After getting a sense of the whole simulation using the 3D animation, they moved on to the contrail evolution view to observe contrail structure formation and whether they merged or dissipated as they evolved (Fig.~\ref{fig:teaser}.E) (A5). There was a larger group structure and multiple smaller groups for each time step. Likewise, in each time step, the smaller groups generally merged into larger groups. To see where these larger structures form in the simulation, they moved to the 3D view and selected the cluster attribute with the MIP shader (Fig.~\ref{fig:contrail_groups}.A) (A4). The experts noticed that those larger groups were far from the jet exit and positioned at the center of the simulation environment. The most senior domain expert confirmed this was a valid scenario, as the more particles move away from the jet exit, the more likely they will turn into ice due to low temperature. To certify if this trend persisted within the simulations, another member was selected, showing a comparable pattern of group formation and evolution (Fig.~\ref{fig:contrail_groups}.B). However, in this simulation run, they saw a contrail group that did not merge to the next time step. The evaluators hypothesized that this group of particles had left the simulation environment, which could mean, in a real-world scenario, that this group had mixed with natural ice clouds. Overall, the domain experts concluded that contrail structure detection could be useful to determine contrail formation and their spread in the environment.

\subsection{Case study 2: Contrail Shape Analysis and Similar Member Detection}
This study aimed to explore contrail shapes and to validate the shape similarity approach visually. These goals were achieved by identifying similar members based on their shapes and I/O parameters (Fig.~\ref{fig:member_similarity}) (A6, A7). The domain experts wished to explore the final time step of multiple simulations, as these time steps correspond to the fully grown contrails. In this study, we examined 10 such simulation runs. The evaluators first used the 3D plume projection view to explore contrail formation and noticed two general trends (A4). One group generated a narrow contrail shape, and the other generated a wide one (Fig.~\ref{fig:member_similarity}.B). 

Based on this finding, they moved on to the similar shape exploration view for further exploration (Fig.~\ref{fig:member_similarity}.A) (A6) of the wide-shaped simulation. They noticed smaller contrail shapes near the jet exit and larger contrail shapes with multiple spikes further away. Subsequently, they looked at similar members based on shape characteristics. They noted that similar simulations also shared the same wide shape with multiple spikes. They hypothesized that several ice particles must have traveled further from the main group, thus generating the spikes. Observing the contrail attributes using the Kiviat diagrams (A6) (Fig.~\ref{fig:member_similarity}.A), the evaluators noticed comparable values among similar members. When considering a narrow-shaped member, they again noticed its similar members shared the same general shape and attributes. Inspecting the shape similarity based on the member I/O parameters (A7), the same members were highlighted, but in a different similarity order. This meant that even though some members had similar shapes, they had different I/O parameters. The wide-shaped members tended to have lower mean temperatures than the narrow-shaped. Surprisingly, some of the narrow-shaped members, even though having lower area values, contained a larger number of particles. The domain experts noted that this view exploration was really important as it could help them to identify suitable parameters to generate different types of simulations, thus achieving optimal parameters for contrail formation.

\subsection{Expert Feedback}
The proposed framework yielded excellent feedback from the domain experts. \textcolor{black}{Given the participatory design process, the feedback was primarily focused on the functionality of the tool.} The 3D plume projection view facilitated the validation of different simulation runs (Fig.~\ref{fig:teaser}.D). The experts confirmed that having pre-computed attributes helped them focus on gaining valuable insights and testing hypotheses. Similarly, they mentioned that the contrail evolution view (Fig.~\ref{fig:teaser}.E) provided the means to observe contrail structure formation and progress, allowing them to assess the mixing of contrail structures with natural ice clouds. In addition, they found the shape similarity exploration view particularly useful (Fig.~\ref{fig:member_similarity}.A)  to generate additional simulation runs and mentioned that, in the future, this would be helpful when finding ideal case scenarios for contrail formation.

\textcolor{black}{The experts noted that the I/O parameter view (Fig.~\ref{fig:teaser}.A-C) helped them identify parameters for contrail formation. Likewise, they found the color glyph helpful when identifying the specific setup for each simulation. Given the number of parameters involved in the analysis, the choice of visualizing only the groups of simulations that differ in the physical or boundary conditions was ``very effective in this respect''.} The multiple-linked paradigm provided an efficient overview, as well as details of the simulation data. The domain experts agreed that the visual front-end’s performance was reasonable, easy to use, and had great applicability in engine simulation research.  

\textcolor{black}{Overall, the experts appreciated the tool developed for the analysis and visualization of contrail data, as well as the specific encodings used: ``The glyph is helpful to identify the setup for each simulation. The visualization of ice crystals, colored with physical features such as ice mass or ice radius, is instructive to characterize the contrail structure. Furthermore, it gives a quick sense of how such a structure depends on the parameters of the simulation. For example, it is very impressive to see through the 3D snapshots reported in Fig.~\ref{fig:member_similarity} how decreasing ambient temperature leads to a much wider contrail, which is physically sound because lower temperature favors vapor condensation into ice. This analysis would have been much harder to perform using simple post-processing without this visualization tool.''}

\section{Discussion}

The credibility of HPC simulations used for environmental predictions to develop public policy, safety procedures and estimate the impact on the environment is of great importance~\cite{verification2002william}. Hence, domain experts need to carefully assess the data quality of these computer-generated simulations to achieve accurate resolutions. Using our system, they were able to validate the data and the models developed; and, in some cases, observed unusual attribute values and anomalies in the distribution of the data. In the end, our work facilitated a blend of computational and human effort to validate the HPC data quality and study contrail formation.

The case studies and the domain expert feedback demonstrate our system’s ability to help identify contrail structures and their evolution over time. Our integrated approach can capture correlations and inconsistencies between input parameters and outputs from multiple simulations. In addition, our approach efficiently supports the analysis of individual time steps and whole simulations by handling one or more members. Through an ACD approach and visual scaffolding~\cite{marai2015visual}, we introduced customized novel encodings to our domain experts (Fig.~\ref{fig:teaser}.B, D), thus enabling them to perform more complex analyses despite low initial visual literacy. For example, through multiple iterations and prototyping, our system introduced customized colored-tile glyphs to show input parameters and boundary conditions of simulation runs. The glyphs scale well with the number of parameters~\cite{aurisano2015bactogenie,luciani2014large}. Despite the large size of the original dataset, this project also supported a detailed investigation of contrail group progressions through a custom tracking graph. Furthermore, our multi-view design supports both a details-first~\cite{details2019luciani} and an overview-first paradigm~\cite{shneiderman1997eyes}, which, as shown by our evaluation, can provide more flexibility in data exploration. \textcolor{black}{The visual front-end follows a standard left-to-right flow, although domain experts can move freely between components.}

With the help of the similar shape exploration view, our evaluators identified similar members based on their contrail shapes or I/O parameters. This is particularly important as domain experts often want to identify input parameter conditions for generating similar simulations. 

Current research in ensemble visualization focuses on analyzing high-dimensional ensemble data through input parameters or simulation outputs. While a few research works deal with both input parameters and simulation outputs, most do not incorporate both into a front-end ~\cite{details2019luciani, cibulski2017super , ovis2014hollt}, or, if they do, they observe the overview of simulation data at a very high level~\cite{makingSense2020dahshan}. Our visualization approach tackles both sides of the problem, considering input parameters and simulation outputs, and contributing to exploratory analysis of individual members.

As ensemble data consists of multiple HPC simulations with different parameter settings, inspecting the huge simulation output data can be overwhelming. Consequently, it is important to have a system that can handle large data efficiently without compromising its overall validity. Besides that, the effectiveness of a web-based visualization tool can be seriously hindered if it takes into account the HPC data without any pre-processing. Our proposed computational back-end is designed to manipulate a high volume of data. We use the volume rendering technique to significantly minimize the data size without losing  important information. \textcolor{black}{The computational back-end uses a number of existing techniques, which we adapt to our domain. Additionally, we introduced techniques for detecting similar shapes and contrail evolution tracking to solve problems specific to this work. Our visual front-end leverages existing visualizations as well as customized visual encodings. Furthermore, even though our proposed approach is based on domain-specific HPC aircraft-engine simulations, it can be generalized to analyze and explore other spatio-temporal CFD data (e.g., analyzing the fluid mixing problem~\cite{maries2012interactive, marai2016visual, monfort2017deep}).}
\textcolor{black}{Likewise, the contrail clustering method can be used to identify similar groups in spatio-temporal data. The contrail evolution view can be used to analyze the temporal progression of other ensemble data, such as in fluid mixing problems. The shape analysis method can also be incorporated into these other domains to define shape characteristics.}
\textcolor{black}{Our color-tile glyph approach works well with a large and varied number of input parameters, and could  be used for other simulation data across domains, as simulation runs commonly feature largely-similar input parameters and boundary conditions.}

There are several limitations to the current design of our system. First, \textcolor{black}{the 3D plume projection view allows domain experts to compare only two time steps of the same or different simulation run(s) while reducing cognitive load. To explore the ensemble data, the experts leveraged instead the I/O Parameters and Similar Shape Exploration views.} Second, to explore an entire simulation, we used an animation feature to provide an overview of the particles and their properties over time. Even though it works well in our case with relatively few time steps, it can hamper longer observations due to its reliance on short-term memory~\cite{munzner2014visualization}. Third, whereas the inherent visual scalability of filaments with the number of attributes shown is limited, filtering operations help alleviate this issue. Finally, our edge-crossing minimization algorithm in the tracking graph works well for a limited number of nodes and links. Future work includes addressing scalability issues, and automatically highlighting unusual attribute values so domain experts can concentrate on these attributes immediately. 

\section{Conclusion}
In this work, we described the activity-centered design of a visual computing framework that supports the analysis of contrails resulting from multiple HPC aircraft engine simulations. We described the application domain data and activities related to defining contrails and contrail-related attributes, quantifying spatial features, and identifying their similarities. We also leveraged a contrail formation criterion and presented a custom algorithm to detect contrail shapes and their characteristics. Additionally, we introduced a novel blend of data mining and interactive visual encodings that links 3D simulation visualization techniques, parameter details of ensemble members, their evolution over time, and shape characteristics in order to explore trends and anomalies within the data, as well as to detect and analyze formation and evolution of contrail. The evaluation of the resulting framework with domain experts shows that this visual computing approach successfully aids in contrail data investigation.

\acknowledgments{
This work and its authors are supported by the US National Science Foundation awards  IIS-2031095, CDSE-1854815, and CNS-1828265, by the Argonne National Laboratory, and partially by the US National Institutes of Health awards NCI R01CA258827 and NLM R01LM012527. We thank all members of the Electronic Visualization Laboratory and our collaborators for their feedback and support during difficult times. \textcolor{black}{We thank the anonymous reviewers for their detailed and constructive feedback and suggestions.}}

\bibliographystyle{abbrv-doi}

\bibliography{template}
\end{document}